\documentclass[12pt, draftclsnofoot, onecolumn]{IEEEtran}
\usepackage{amssymb}
\usepackage{color}
\usepackage{float}
\usepackage{tabularx,colortbl}
\usepackage{setspace}
\usepackage{cite,graphicx,amsmath,psfrag,bm}
\usepackage{subfigure}
\usepackage[usenames,dvipsnames]{xcolor}
\usepackage{mathabx}
\usepackage{cancel}
\usepackage{epstopdf}
\usepackage[process=auto]{pstool}
\usepackage{caption}
\usepackage[normalem]{ulem}
\usepackage{tablefootnote}
\usepackage{multicol}

\usepackage{mdframed}
\usepackage{tikz}

\graphicspath{{Figs/}}

\newcommand{\figref}{Fig.~\ref}

\DeclareMathOperator{\rect}{rect}

\begin{document}

\title{Time-Reversal Routing for Dispersion Code Multiple Access (DCMA) Communications}

\author{Lianfeng Zou,~\IEEEmembership{Student member,~IEEE,}
        Christophe Caloz,~\IEEEmembership{Fellow,~IEEE}
\thanks{L. Zou (lianfengzou@gmail.com) and C. Caloz are with Electrical Engineering department of \'Ecole Polytechnique de Montr\'{e}al, Montr\'{e}al, Qu\'{e}bec, H3T1J4, Canada.}%
}
\markboth{IEEE Wireless Communications Letters ,~Vol.~, No.~, Month~2017}%
{Shell \MakeLowercase{\textit{et al.}}: Bare Demo of IEEEtran.cls for Journals}

\maketitle

\begin{abstract}
We present the modeling and characterization of a time-reversal routing dispersion code multiple access (TR-DCMA) system. We show that this system maintains the low complexity advantage of DCMA transceivers while offering dynamic adaptivity for practial communication scenarios. We first derive the mathematical model and explain operation principles of the system, and then characterize its interference, signal to interference ratio, and bit error probability characteristics.
\end{abstract}


\IEEEpeerreviewmaketitle

\section{Introduction}

Real-time Analog Signal Processing (R-ASP), a new paradigm for future millimeter-wave and terahertz high-speed wireless communications, consists in processing high-frequency ultrawide-band RF signals in real time using dispersion-engineering electromagnetic components called ``phasers"~\cite{Jour:2013_MwMag_Caloz,JOUR:2015_TMTT_Gupta}. A phaser is an analog processor that ideally provides a flat magnitude response and an application-specific group delay response. R-ASP applications include spectrum analysis~\cite{JOUR:2003_TMTT_Laso}, spectrum sniffing~\cite{JOUR:2012_MWCL_Nikfal}, time-stretching based sampling enhancement~\cite{JOUR:2007_TMTT_Schwartz, JOUR:2012_TMTT_Xiang,JOUR:2011_TMTT_Nikfal}, time reversal~\cite{CONF:2008_IRWS_Schwartz}, chipless RFID~\cite{JOUR:2011_AWPL_Gupta}, communication SNR enhancement~\cite{JOUR:2014_MWCL_Nikfal} and dispersion code multiple access (DCMA) wireless communication
~\cite{CONF:2015_APS_Gupta,JOUR:2017_arxiv_zou}.

DCMA wireless communication is a promising R-ASP technology, where each access point is assigned a distinct dispersion code, or a specified group delay function provided by a phaser, for multiple access with low transceiver complexity and negligible processing latency following from the analog nature of R-ASP~\cite{JOUR:2017_arxiv_zou}. However, DCMA systems reported to date can only route signals between fixed communication pairs. For dynamic routing between arbitrary pairs, an adaptivity strategy must be introduced. One solution may be to use phasers that reconfigure in real time to match the group delay profiles between arbitrary access point pairs~\cite{JOUR:2017_TMTT_LZOU}, but the technology for such phasers is complex and still at an early development stage. For this reason, we propose a routing station, or router, where the routing may be efficiently performed by using time reversal~\cite{CONF:2015_ISAP_LZOU,JOUR:1992_TUFFC_Fink, JOUR:2004_PhysRevLett_FINK}.

\section{System Description}

\begin{figure*}[h!t]
   \centering
   \includegraphics[width=1.0\linewidth]{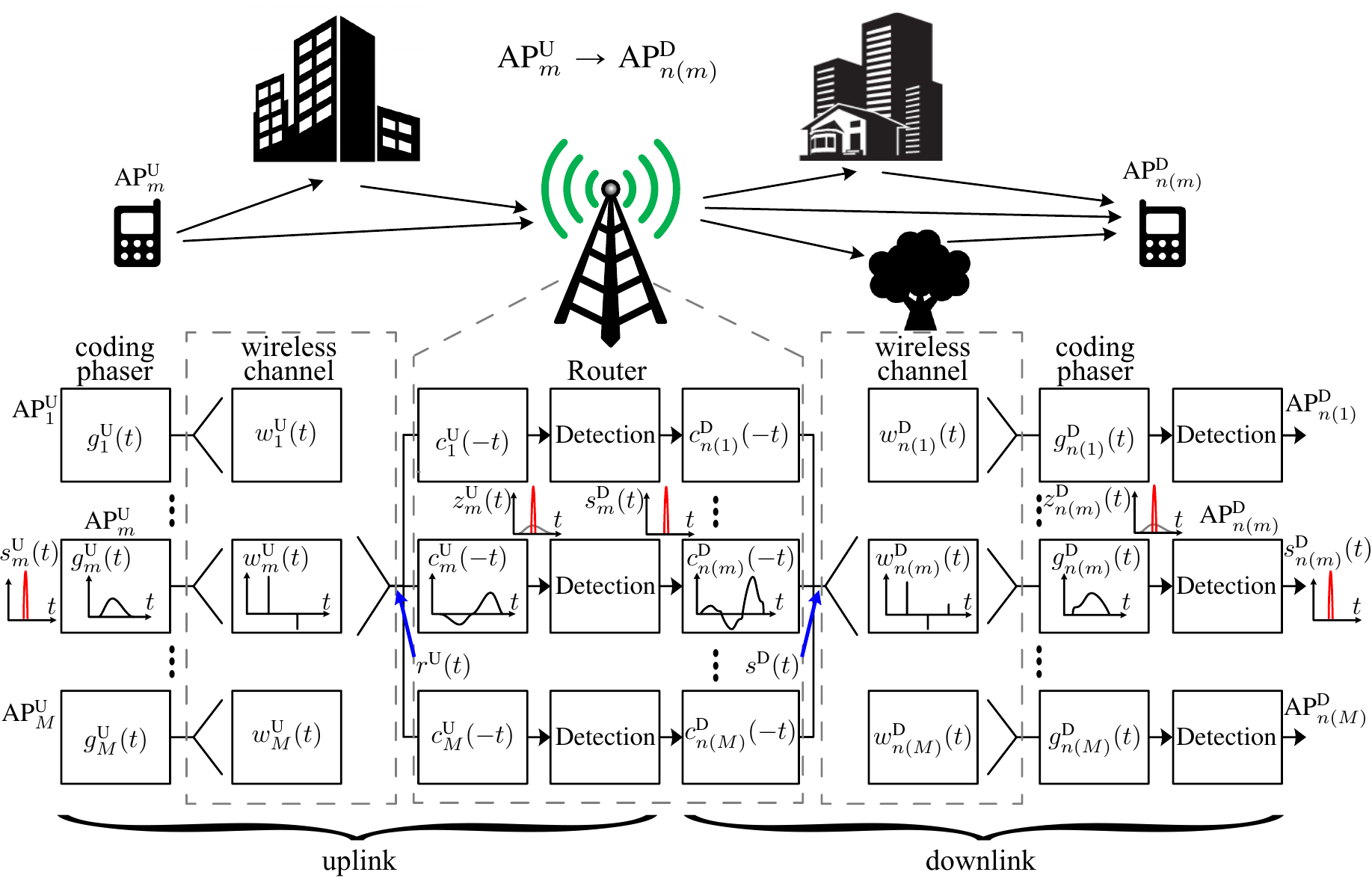}
   \caption{Diagrammatic representation of the proposed TR-DCMA system.}
   \label{FIG:SYSTEM}
\end{figure*}

Figure~\ref{FIG:SYSTEM} shows a diagrammatic representation of the proposed Time-Reversal Dispersion Code Multiple Access (TR-DCMA) routing system with $2M$ access points (AP) and the router with endowed with time reversal capability. Uplink AP$^\text{U}_m,\, m\in\{1,\ldots, M\}$, communicates with downlink AP$^\text{D}_{n(m)}$, $n(m)\in\{1,\ldots,M\}$, via the router, where $n(m)$ is a function of $m$ corresponding to the desired routing link from access point $m$ to access point $n$, with $n(m_1) \neq n(m_2)$ for $m_1\neq m_2$.

For multiple access purpose, AP$^\text{U/D}_k$ is assigned a specific dispersion code, which is the group delay function $\tau_k^\text{U/D}(\omega)$, provided by the coding phaser~\cite{Jour:2013_MwMag_Caloz} that is incorporated in the AP system before/after the antenna. The phaser impulse response $g_k^\text{U/D}(t)$ is found by inverse Fourier transforming ($\mathcal{F}^{-1}$) the transfer function $G_k^\text{U/D}(\omega)$ as
 \begin{subequations}\label{EQ:PRC_SYS_PHSRIR}
  \begin{equation}\label{EQ:PRC_SYS_PHSRIR_FT}
    g^\text{U/D}_k(t) =\mathcal{F}^{-1}\left[G^\text{U/D}_k(\omega)\right]
    = \mathcal{F}^{-1}\left[\rect\left(\dfrac{\omega-\omega_0}{\Delta\omega}\right)e^{j\phi^\text{U/D}_k(\omega)}\right],
\end{equation}
where
\begin{equation}\label{EQ:PRC_SYS_PHSRIR_PHS}
 \phi^\text{U/D}_k(\omega) = -\int_{\omega_0-\Delta\omega/2}^{\omega}\tau^\text{U/D}_k (\omega')\,d\omega',
\end{equation}
\end{subequations}
and $\tau^\text{U/D}_k(\omega)$ are the phaser transfer phase and group delay (dispersion code), respectively, and $\omega_0 = 2\pi f_0$, $\Delta\omega = 2\pi\Delta f$ are the center frequency and bandwidth, respectively. The \textit{wireless} channel between the AP (after/before the phaser) and the router, denoted as $w^\text{U/D}_k(t)$, naturally includes the AP and the router antenna impulse responses in the communication direction and typically exhibits multipath fading~\cite{BK:2011_MOLISCH_WIRELESSCOMM}.

\section{Modeling}
\subsection{Calibration Phase}\label{SEC:PRC_PHASE1}

During this phase, the $2M$ APs sequentially send a known beacon signal, $s^\text{B}(t)$, to the router. The router receives for AP$^\text{U/D}_k$ the signal
\begin{subequations}
\begin{equation}\label{EQ:PRC_MEASBEACON}
\begin{split}
  r^\text{B,U/D}_k(t) & = \left[(s^\text{B}\ast g^\text{U/D}_k)\ast w^\text{U/D}_k\right](t)\\
  & = \left(s^\text{B} \ast c^\text{U/D}_k\right)(t),
\end{split}
\end{equation}
where
\begin{equation}\label{EQ:PRC_SYS_CHAN}
 c^\text{U/D}_k(t) = (g^\text{U/D}_k \ast w^\text{U/D}_k)(t),
\end{equation}
\end{subequations}
is the overall channel impulse response, corresponding to the convolution (``$\ast$") of the corresponding \textit{guided-wave} channel (coding phaser) and \textit{wireless} channel impulse response.

Since $s^\text{B}(t)$ is known, $c^\text{U/D}_k(t)$ can be determined from~\eqref{EQ:PRC_MEASBEACON} by the router. This may be done either digitally or analogically. In the former case, the measured signal $r^\text{U/D}_k(t)$ is stored, then numerically deconvolved and flipped, i.e. $c^\text{U/D}_k(t)\rightarrow c^\text{U/D}_k(-t)$, and finally reconverted to the analog domain. In the latter case, $r^\text{U/D}_k(t)$ is immediately deconvolved by the (known) time-reversed version of $s^\text{B}(t)$, $s^\text{B}(-t)$, using a real-time convolver~\cite{JOUR:2016_JLT_JZhang}, yielding $c^\text{U/D}_k(t)$, which is itself time-reversed by a real-time time reverser~\cite{CONF:2008_IRWS_Schwartz} into $c^\text{U/D}_k(-t)$.

\subsection{Communication Phase}
\subsubsection{Uplink Transmission }\label{SEC:PRC_PHASE2}

Assume the worst-case scenario where the $M$ uplink APs are sending their signals at the same time. Denoting $s^\text{U}_m(t)$ the signal sent from AP$^\text{U}_m$, the signal received by the router is
\begin{equation}\label{EQ:PRC_SYS_RTRECSIG}
  r^\text{U}(t) = \sum_{m=1}^M \alpha^\text{U}_ms^\text{U}_m(t)\ast c^\text{U}_m(t),
\end{equation}
where $\alpha^\text{U}_m>0$ is the sent signal magnitude. The decoding of the signal from AP$^\text{U}_m$ at the router consists in convolving $r^\text{U}(t)$ with the time-reversed version of the corresponding channel impulse response $c^\text{U}_m(-t)$ constructed in the calibration phase [Sec.~\eqref{SEC:PRC_PHASE1}]. Thus,
\begin{subequations}\label{EQ:PRC_ROUTERDECSIG}
\begin{equation}\label{EQ:PRC_ROUTERDECSIG1}
  z^\text{U}_m(t)  = r^\text{U}(t)\ast c^\text{U}_m(-t)
   = \tilde{s}^\text{U}_m(t) + x^\text{U}_m(t),
\end{equation}
where
\begin{equation}\label{EQ:PRC_ROUTERDECSIG2}
  \tilde{s}^\text{U}_m(t) = \alpha^\text{U}_ms^\text{U}_m(t) \ast  c^\text{U}_m(t) \ast c^\text{U}_m(-t)
  \approx\alpha^\text{U}_ms^\text{U}_m(t),
\end{equation}
is an approximation of the desired signal, $s^\text{U}_m(t)$, the approximation (rather equality) being due to the finite calibration time in~\eqref{EQ:PRC_SYS_CHAN}\footnote{If the calibration time were infinite, then we would have an equality from the identity $c^\text{U}_m(t) \ast c^\text{U}_m(-t)=\delta(t)$. In practice, $c^\text{U}_m(t)$ in~\eqref{EQ:PRC_ROUTERDECSIG2} is a \emph{truncated} version of the ideal $c^\text{U}_m(t)$ function.}, and
\begin{equation}\label{EQ:PRC_ROUTERDECSIG3}
  x^\text{U}_m(t) = c^\text{U}_m(-t)\ast \sum_{\substack{k=1\\k\neq m}}^M \alpha^\text{U}_ks^\text{U}_k(t)\ast  c^\text{U}_k(t),
\end{equation}
\end{subequations}
is a distortion signal called multiple-access interference (MAI).

\subsubsection{Router Detection}

At this point, the uplink signal $z^\text{U}_m(t)$ in~\eqref{EQ:PRC_ROUTERDECSIG}, including the desired information $\tilde{s}^\text{U}_m(t)$ and interference from the other channels $x^\text{U}_m(t)$, is passed through a threshold detector in the router (Fig.~\ref{FIG:SYSTEM}), which transforms it into the signal $s^\text{D}_m(t)$.

\subsubsection{Downlink Transmission}

In the downlink transmission process, the signal $s^\text{D}_m(t)$ is to be routed to AP$^\text{D}_{n(m)}$, the desired corresponding access point, that generally varies in time. For this purpose, it is first predistorted by convolution with the time-reversed version of the corresponding downlink channel impulse response, $c^\text{D}_{n(m)}(-t)$. Then, the $M$ predistorted signals are combined and sent by the antenna of the router as
\begin{equation}\label{PRC_ROUTERSENTSIG}
  s^\text{D}(t) = \sum_{m=1}^{M}\alpha^\text{D}_m s^\text{D}_m(t) \ast c^\text{D}_{n(m)}(-t),~\alpha^\text{D}_m>0.
\end{equation}

After passing the wireless channel $w^\text{D}_{n(m)}(t)$, this signal is decoded by phaser $g^\text{D}_{n(m)}(t)$ as
\begin{subequations}\label{EQ:PRC_RXDECSIG}
\begin{equation}\label{EQ:PRC_RXDECSIG1}
\begin{split}
  z^\text{D}_{n(m)}(t) & = s^\text{D}(t)\ast w^\text{D}_{n(m)}(t)\ast g^\text{D}_{n(m)}(t) \\
  &= s^\text{D}(t)\ast c^\text{D}_{n(m)}(t)
   = \tilde{s}^\text{D}_{n(m)}(t) + x^\text{D}_{n(m)}(t),
   \end{split}
\end{equation}
where
\begin{equation}\label{EQ:PRC_RXDECSIG2}
  \tilde{s}^\text{D}_{n(m)}(t) = \alpha^\text{D}_m s^\text{D}_m(t) \ast  c^\text{D}_{n(m)}(-t) \ast c^\text{D}_{n(m)}(t)
  \approx\alpha^\text{D}_m s^\text{D}_m(t)
\end{equation}
and
\begin{equation}\label{EQ:PRC_RXDECSIG3}
  x^\text{D}_{n(m)}(t) =c^\text{D}_{n(m)}(t) \ast \sum_{\substack{k=1\\k\neq {m}}}^M \alpha^\text{D}_k s^\text{D}_k(t)\ast c^\text{D}_{n(k)}(-t).
\end{equation}
\end{subequations}

The following threshold detection (Fig.~\ref{FIG:SYSTEM}) yields $s^\text{D}_{n(m)}(t)$. Communication is naturally successful when the detected downlink signal is identical to the transmitted uplink signal, i.e. $s^\text{D}_{n(m)}(t) = s^\text{D}_m(t)=s^\text{U}_m(t)$.

\section{System Characterization}\label{SEC:MAISIR}

This section characterizes the proposed time-reversal routing DCMA system in terms of MAI, signal to interference ratio (SIR) and bit error probability (BEP) for the case of On-Off Keying (OOK) modulation and  Chebyshev dispersion coding. Note that, since uplink and downlink signals are described by mathematical expressions, Eqs.~\eqref{EQ:PRC_ROUTERDECSIG} and~\eqref{EQ:PRC_RXDECSIG}, of the same form, we shall consider here only the uplink case, the downlink and overall transmission being immediately deducible from it.

\subsection{Modulation and Coding}

Assuming OOK modulation, the transmitted signal is the pulse train
\begin{equation}\label{EQ:CHAR_UL_SIG}
 s^\text{U}_m(t)  = \sum_{\ell}d^\text{U}_{m\ell}\delta(t-\ell T_\text{b} - t^\text{U}_m),
\end{equation}
where $d^\text{U}_{m\ell}=1$ or $0$ is the $\ell^\text{th}$ base-band bit, $\delta(\cdot)$ is the Dirac function, $T_\text{b}$ is bit period and $t^\text{U}_m$ is a random time offset.

Following~\cite{JOUR:2017_arxiv_zou}, we choose odd Chebyshev dispersion coding [$\tau^\text{U}_m(\omega)$] for AP$^\text{U}_m,\,\forall\,m$, corresponding to
\begin{equation}\label{EQ:CHAR_UL_CHEBYCODES}
  \tau^\text{U}_m(\omega) = \tau_0 + \dfrac{\Delta\tau}{2}T_{i(m)}\left(\dfrac{\omega-\omega_0}{\Delta\omega/2}\right),
\end{equation}
where $\Delta\tau$ is group delay swing over the band $\Delta\omega$, $T_{i(m)}$ is $i(m)^\text{th}$ order Chebyshev polynomial of the first kind, and where we define $T_{-i(m)}=-T_{i(m)}$ for $i(m)>0$. The code set of the $M$ uplink access points may then be written
\begin{equation}\label{EQ:CHAR_UL_CODESET}
\begin{split}
  &\mathbf{C} = \{i(1),\ldots, i(m),\ldots, i(M)\},~i(m)~\text{odd} \text{ and } i(m)\geq3.
\end{split}
\end{equation}
In the forthcoming computations, we consider the CM3 type [4--10~m non-line-of-sight (non-LOS)] indoor multipath channel~\cite{REPORT:2003_Foerster_UWBCH} for $w^\text{U}_m(t)$.

\subsection{MAI Probability Density Function}\label{SEC:MAI}

In~\cite{JOUR:2017_arxiv_zou}, we have shown that in a LOS wireless channel, the MAI corresponding to all-odd Chebyshev dispersion coding~\eqref{EQ:CHAR_UL_CODESET} follows a normal distribution. We shall show here that the same is true for non-LOS.

\begin{figure}[h!t]
   \centering
   \includegraphics[width=0.95\linewidth]{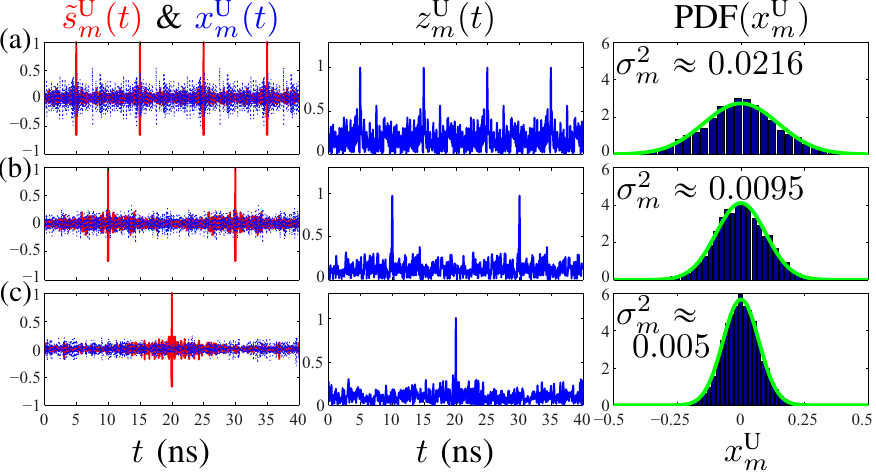}
   \caption{Uplink simulation results in the worst-case interference scenario, where $d_{m\ell}^\text{U}=1$, $\forall m,\ell$, in~\eqref{EQ:CHAR_UL_SIG}
    for $M=5$ TR-DCMA with $\Delta f = 10$~GHz, $\Delta\tau = 10$~ns, coding $\mathbf{C}=\{3,-3,5,-5,7\}$ and identical energy $\alpha^\text{U}_m=$ const. in~\eqref{EQ:PRC_SYS_RTRECSIG}. All the results are normalized as follows: for each~$m$, $\alpha_m^\text{U}$ is set such that $|\tilde{s}_m^\text{U}(t)|_\text{max}=1$ in~\eqref{EQ:PRC_ROUTERDECSIG2} and $x_m^\text{U}(t)$ is divided by that $\alpha_m^\text{U}$ in~\eqref{EQ:PRC_ROUTERDECSIG3}. First column: desired signal, $\tilde{s}^\text{U}_m(t)$ (red-solid curve), and MAI, $x^\text{U}_m(t)$ (blue-dotted curve), computed using~\eqref{EQ:PRC_ROUTERDECSIG2} and~\eqref{EQ:PRC_ROUTERDECSIG3}, respectively. Second column: total encoded signal, $z^\text{U}_m(t)$, computed using~\eqref{EQ:PRC_ROUTERDECSIG1}. Third column: probability density function (PDF) of the MAI values, obtained by counting the occurrences of the sample values (blue stripes) and compared against the normal distribution PDF (green curve) [Eq.~\eqref{EQ:CHAR_UL_MAIDIST} with mean $\mu_m=0$, $\forall m$, and variance $\sigma_m^2$ in~\eqref{EQ:CHAR_UL_MAIVAR}. (a)~$T_\text{b}=\Delta\tau$, (b)~$T_\text{b}=2\Delta\tau$, and (c)~$T_\text{b}=4\Delta\tau$.}
   \label{FIG:MAI}
\end{figure}

Figure~\ref{FIG:MAI} plots uplink simulation results of an $M=5$ TR-DCMA system for three different bit periods ($T_\text{b}$) in the worst-case interference scenario where all the transmitters continuously send the bit '1'. As expected, the interference (MAI) floor [$x^\text{U}_m(t)$] decreases with  increasing $T_\text{b}$ due to decreasing MAI interference. The probability density function (PDF) of $x^\text{U}_m(t)$ are found (third column in the figure) to closely follow the normal distribution PDF
\begin{subequations}
\begin{equation}\label{EQ:CHAR_UL_MAIDIST}
 \text{PDF}(x^\text{U}_m) = \dfrac{1}{\sqrt{2\pi\sigma^2}}\exp{\left[-\dfrac{(x^\text{U}_m-\mu_m)^2}{2\sigma_m^2}\right]},
\end{equation}
where $\mu_m$ is the mean of $x^\text{U}_m$, which is $0$ due to the symmetric-bipolar nature of MAI, and $\sigma_m^2$ is the variance,
\begin{equation}\label{EQ:CHAR_UL_MAIVAR}
   \sigma_m^2 = \dfrac{1}{T_\text{b}}\int_{T_\text{b}} \left|x^\text{U}_m(t)-\mu^\text{U}_m\right|^2\,dt = \dfrac{1}{T_\text{b}}\int_{T_\text{b}}  \left|x^\text{U}_m(t)\right|^2\,dt,
\end{equation}
\end{subequations}
which is equivalent to the MAI average power over one bit. In a realistic scenario, where bits '1' and '0' alternate in the wireless channel, the interference would naturally be less, leading to smaller $\sigma_m^2$ values. In the forthcoming results, the same worst-case scenario has been assumed for the PDF, and practical results would then be better than what will be shown.

\subsection{Statistical and Analytical SIR}\label{SEC:CHAR}

The SIR may be statistically found by taking the ratio of $|\tilde{s}_m^\text{U}(t)|_\text{max}$ to the MAI variance given by~\eqref{EQ:CHAR_UL_MAIVAR}, using the normalization indicated in the caption of Fig.~\ref{FIG:MAI}, which yields
\begin{equation}\label{EQ:CHAR_UL_SIR1}
{\text{SIR}_m^\text{U}}'=\dfrac{1}{\sigma_m^2}.
\end{equation}

This quantity can also be obtained analytically as~\cite{JOUR:2017_arxiv_zou}
\begin{subequations}\label{EQ:CHAR_UL_SIR2}
  \begin{equation}\label{EQ:CHAR_UL_SIR21}
  \text{SIR}_m^\text{U} = \dfrac{2\Delta f T_\text{b}}{\overline{\alpha_m^2}(M-1)},
\end{equation}
where
  \begin{equation}\label{EQ:CHAR_UL_SIR22}
  \overline{\alpha_m^2}= \dfrac{1}{M-1}\sum_{\substack{k=1\\k\neq {m}}}^M \left(\dfrac{\alpha^\text{U}_k}{\alpha^\text{U}_m}\right)^2
\end{equation}
\end{subequations}
is the mean of the normalized MAI energies\footnote{In~\eqref{EQ:CHAR_UL_SIR22}, $k=m$ is excluded from the sum as it corresponds to the signal.}.

\begin{figure}[h!t]
   \centering
   \includegraphics[width=0.6\linewidth]{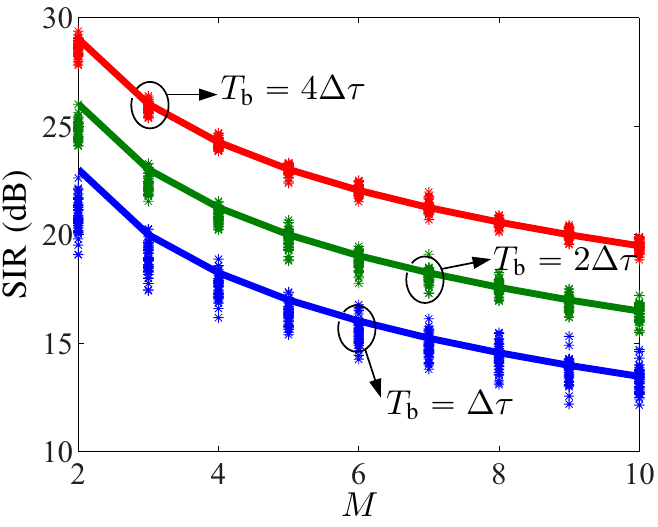}
   \caption{SIR versus the number of transmitters ($M$) with $\Delta f = 10$~GHz, $\Delta\tau = 10$~ns, coding $\mathbf{C} = \{3, -3, 5, -5,\ldots\}$, identical energy ($\alpha^\text{U}_m=$ const. $\forall\,m$) in~\eqref{EQ:PRC_SYS_RTRECSIG}, and different $T_\text{b}$. Solid curves: Eq.~\eqref{EQ:CHAR_UL_SIR2}, `*' markers: Eq.~\eqref{EQ:CHAR_UL_SIR1} with~\eqref{EQ:CHAR_UL_MAIVAR} and~\eqref{EQ:PRC_ROUTERDECSIG3} for $500$ bits.  }
   \label{FIG:SIR}
\end{figure}

Figure~\ref{FIG:SIR} compares the analytical [Eq.~\eqref{EQ:CHAR_UL_SIR2}] and statistical [Eq.~\eqref{EQ:CHAR_UL_SIR1}] SIRs. Good agreement is observed, with deviation smaller than $2$~dB. Therefore, we will directly use~\eqref{EQ:CHAR_UL_SIR2} to avoid statistical testing over many bits in the remainder of the paper.

The downlink MAI also follows normal distribution, and the corresponding SIR$^\text{D}_{n(m)}$ is also approximated by~\eqref{EQ:CHAR_UL_SIR2} with $\alpha^\text{U}_k$ and $\alpha^\text{U}_m$ replaced by $\alpha^\text{D}_{n(k)}$ and $\alpha^\text{D}_{n(m)}$.

\subsection{Overall BEP Performance}

The BEP for MAI with normal distribution is~\cite{JOUR:2017_arxiv_zou}
\begin{equation}\label{EQ:BEP_BEPUD}
  \text{BEP}^\text{U}_m   = \dfrac{1}{\sqrt{2\pi}}\int_{\sqrt{\text{SIR}^\text{U}_m}/2}^{\infty}\exp\left(-\dfrac{x^2}{2}\right)\,dx,
\end{equation}
where $\text{SIR}^\text{U}_m$ is given by~\eqref{EQ:CHAR_UL_SIR2}. The downlink BEP$^\text{D}_{n(m)}$ is found by replacing SIR$^\text{U}_m$ in~\eqref{EQ:BEP_BEPUD} with SIR$^\text{D}_{n(m)}$.

Communication is overall successful if both the uplink and downlink transmissions are successful, corresponding to the overall BEP
\begin{equation}\label{EQ:BEP_BEPO}
\begin{split}
  \text{BEP}_m &= 1- \left(1-\text{BEP}^\text{U}_m\right)\left(1-\text{BEP}^\text{D}_{n(m)}\right)\\
  & = \text{BEP}^\text{U}_m + \text{BEP}^\text{D}_{n(m)} - \text{BEP}^\text{U}_m\text{BEP}^\text{D}_{n(m)}\\
  & \approx \text{BEP}^\text{U}_m + \text{BEP}^\text{D}_{n(m)}.
\end{split}
\end{equation}

\begin{figure}[h!t]
   \centering
   \includegraphics[width=0.6\linewidth]{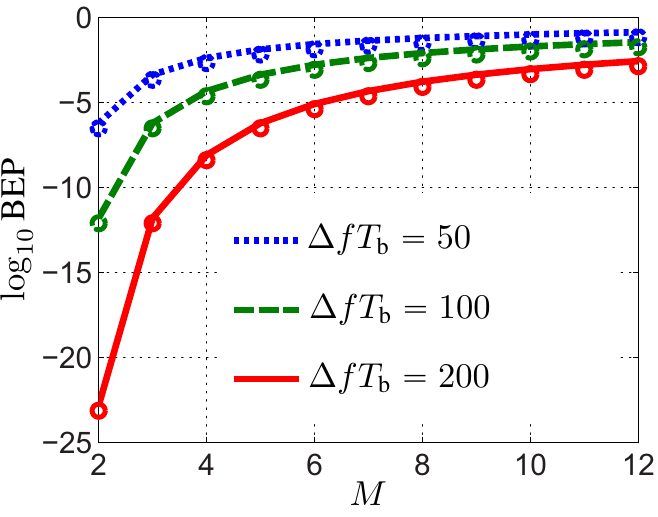}
   \caption{BEP versus the number of simultaneous communication links ($M$) in the TR-DCMA system in~\figref{FIG:SYSTEM} for APs with identical energy ($\alpha^\text{U}_m=\alpha^\text{D}_{n(m)}=$ const. $\forall\,m$), computed using~\eqref{EQ:BEP_BEPO} (curves), and compared against the BEP of the corresponding DCMA system without time-reversal routing~\cite{JOUR:2017_arxiv_zou} (circles), for different $\Delta f T_\text{b}$ values.   }
   \label{FIG:BEP}
\end{figure}

Figure~\ref{FIG:BEP} plots the BEP (same for all $m$'s) of the TR-DCMA system for APs with identical energy, and compared against that of the corresponding DCMA system without time-reversal routing. Due to the two-step (uplink and downlink) transmission phases, the BEP is approximately doubled, or degraded by an order of $\log_{10}2\approx0.3$. This graph shows that the BEP is not affect by the dynamic TR routing.

\section{Conclusion}\label{SEC:CONCL}
A TR-DCMA routing system has been presented and characterized in terms of MAI statistical distributions, SIR, and BEP. The system may find its applications in dynamic wireless communications requiring low-complexity transceivers and negligible latency.
\bibliographystyle{IEEEtran}
\bibliography{IEEEabrv,DCMA_REF}

\begin{thebibliography}{10}
\providecommand{\url}[1]{#1}
\csname url@samestyle\endcsname
\providecommand{\newblock}{\relax}
\providecommand{\bibinfo}[2]{#2}
\providecommand{\BIBentrySTDinterwordspacing}{\spaceskip=0pt\relax}
\providecommand{\BIBentryALTinterwordstretchfactor}{4}
\providecommand{\BIBentryALTinterwordspacing}{\spaceskip=\fontdimen2\font plus
\BIBentryALTinterwordstretchfactor\fontdimen3\font minus
  \fontdimen4\font\relax}
\providecommand{\BIBforeignlanguage}[2]{{%
\expandafter\ifx\csname l@#1\endcsname\relax
\typeout{** WARNING: IEEEtran.bst: No hyphenation pattern has been}%
\typeout{** loaded for the language `#1'. Using the pattern for}%
\typeout{** the default language instead.}%
\else
\language=\csname l@#1\endcsname
\fi
#2}}
\providecommand{\BIBdecl}{\relax}
\BIBdecl

\bibitem{Jour:2013_MwMag_Caloz}
C.~Caloz, S.~Gupta, Q.~Zhang, and B.~Nikfal, ``Analog signal processing: A
  possible alternative or complement to dominantly digital radio schemes,''
  \emph{IEEE Microw. Mag.}, vol.~14, no.~6, pp. 87 -- 103, Sep. 2013.

\bibitem{JOUR:2015_TMTT_Gupta}
S.~Gupta, Q.~Zhang, L.~Zou, L.~Jiang, and C.~Caloz, ``Generalized coupled-line
  all-pass phasers,'' \emph{IEEE Trans. Microw. Theory Techn.}, vol.~63, no.~3,
  pp. 1 -- 12, Mar. 2015.

\bibitem{JOUR:2003_TMTT_Laso}
M.~A.~G. Laso, T.~Lopetegi, M.~J. Erro, D.~Benito, M.~J. Garde, M.~A. Muriel,
  M.~Sorolla, and M.~Guglielmi, ``Real-time spectrum analysis in microstrip
  technology,'' \emph{IEEE Trans. Microw. Theory Techn.}, vol.~51, no.~3, pp.
  705 -- 717, Mar. 2003.

\bibitem{JOUR:2012_MWCL_Nikfal}
B.~Nikfal, D.~Badiere, M.~Repeta, B.~Deforge, S.~Gupta, and C.~Caloz,
  ``Distortion-less real-time spectrum sniffing based on a stepped group-delay
  phaser,'' \emph{IEEE Microw. Wireless Compon. Lett.}, vol.~22, no.~11, pp.
  601 -- 603, Nov. 2012.

\bibitem{JOUR:2007_TMTT_Schwartz}
J.~D. Schwartz, J.~Aza\~{n}a, and D.~V. Plant, ``A fully electronic system for
  the time magnification of ultra-wideband signals,'' \emph{IEEE Trans. Microw.
  Theory Techn.}, vol.~55, no.~2, pp. 327 -- 334, Feb. 2007.

\bibitem{JOUR:2012_TMTT_Xiang}
B.~Xiang, A.~Kopa, Z.~Fu, and A.~B. Apsel, ``Theoretical analysis and practical
  considerations for the integrated time-stretching system using dispersive
  delay line ({DDL}),'' \emph{IEEE Transactions on Microwave Theory and
  Techniques}, vol.~60, no.~11, pp. 3449 -- 3457, Nov. 2012.

\bibitem{JOUR:2011_TMTT_Nikfal}
B.~Nikfal, S.~Gupta, and C.~Caloz, ``Increased group delay slope loop system
  for enhanced-resolution analog signal processing,'' \emph{IEEE Trans. Microw.
  Theory Techn.}, vol.~59, no.~6, pp. 1622 -- 1628, Jun. 2011.

\bibitem{CONF:2008_IRWS_Schwartz}
J.~D. Schwartz, J.~Aza\~{n}a, and D.~V. Plant, ``An electronic temporal imaging
  system for compression and reversal of arbitrary {UWB} waveforms,'' in
  \emph{Proc. IEEE Radio and Wireless Symp.}, Orlando, FL. U.S., Jan. 2008, pp.
  487 -- 490.

\bibitem{JOUR:2011_AWPL_Gupta}
S.~Gupta, B.~Nikfal, and C.~Caloz, ``Chipless {RFID} system based on group
  delay engineered dispersive delay structures,'' \emph{IEEE Antennas and
  Wireless Propag. Lett.}, vol.~10, pp. 1366 -- 1368, Oct. 2011.

\bibitem{JOUR:2014_MWCL_Nikfal}
B.~Nikfal, Q.~Zhang, and C.~Caloz, ``Enhanced-{SNR} impulse radio transceiver
  based on phasers,'' \emph{IEEE Microw. Wireless Compon. Lett.}, vol.~24,
  no.~11, pp. 778 -- 780, Nov. 2014.

\bibitem{CONF:2015_APS_Gupta}
S.~Gupta, L.~Zou, M.~A. Salem, and C.~Caloz, ``Bit-error-rate ({BER})
  performance in dispersion code multiple access ({DCMA}),'' in \emph{Proc.
  IEEE Int. Symp. on Antennas Propag.}, Jul. 2015, pp. 1015 -- 1016.

\bibitem{JOUR:2017_arxiv_zou}
\BIBentryALTinterwordspacing
L.~Zou, S.~Gupta, and C.~Caloz, ``Real-time dispersion code multiple access
  {(DCMA)} for high-speed wireless communications,'' \emph{arXiv Information
  Theory}, vol. abs/1703.10516, 2017. [Online]. Available:
  \url{http://arxiv.org/abs/1703.10516}
\BIBentrySTDinterwordspacing

\bibitem{JOUR:2017_TMTT_LZOU}
------, ``Loss-gain equalized reconfigurable c-section analog signal
  processor,'' \emph{IEEE Trans. Microw. Theory Techn.}, vol.~65, no.~2, pp.
  555 -- 564, Feb. 2017.

\bibitem{CONF:2015_ISAP_LZOU}
------, ``Time-reversal based routing in dispersion code multiple access
  (dcma),'' in \emph{IEEE Int. Symp. Antennas Propag. (ISAP)}, Hobart,
  Australia, Nov. 2015, pp. 366 -- 369.

\bibitem{JOUR:1992_TUFFC_Fink}
M.~Fink, ``Time reversal of ultrasonic fields-part i: Basic principles,''
  \emph{IEEE Trans. Ultrason., Ferroelect., Freq. Control}, vol.~39, no.~5, pp.
  555 -- 566, May 1992.

\bibitem{JOUR:2004_PhysRevLett_FINK}
G.~Lerosey, J.~de~Rosny, A.~Tourin, A.~Derode, G.~Montaldo, and M.~Fink, ``Time
  reversal of electromagnetic waves,'' \emph{Phys. Rev. Lett.}, vol.~92, p.
  193904, May 2004.

\bibitem{BK:2011_MOLISCH_WIRELESSCOMM}
A.~F. Molisch, \emph{Wireless Communications, 2nd Ed.}\hskip 1em plus 0.5em
  minus 0.4em\relax John Wiley \& Sons, Inc., 2011.

\bibitem{JOUR:2016_JLT_JZhang}
J.~Zhang and J.~Yao, ``Photonic-assisted microwave temporal convolution,''
  \emph{J. Lightw. Technol.}, vol.~34, no.~20, pp. 4652--4657, Oct 2016.

\bibitem{REPORT:2003_Foerster_UWBCH}
J.~R. Foerster, M.~Pendergrass, and A.~F. Molisch, ``A channel model for ultra
  wideband indoor communication,'' Mitsubishi Electric Research Laboratories,
  Inc., Tech. Rep., Oct. 2003.

\end{thebibliography}

\end{document}